\documentclass[runningheads]{llncs}

\usepackage{hyperref}
\usepackage{xcolor}

\let\svthefootnote\thefootnote
\newcommand\freefootnote[1]{%
  \let\thefootnote\relax%
  \footnotetext{#1}%
  \let\thefootnote\svthefootnote%
}

\usepackage[listings,skins,breakable,xparse]{tcolorbox}

\lstdefinelanguage{turtle}
{
    columns=fullflexible,
    keywordstyle=\color{red},
    morekeywords={@prefix,@base,@forSome,@forAll,@keywords},
    morecomment=[l]{\#},
    tabsize=4,
    alsoletter={-?},
    basicstyle=\ttfamily\color{black},
    morestring=[b][\color{black}]\"
}

\usepackage{chngcntr}
\AtBeginDocument{\counterwithout{lstlisting}{chapter}}

\begin{document}
\title{PREC: semantic translation of property graphs}
\titlerunning{PREC}

\author{Julian Bruyat\inst{1} \and Pierre-Antoine Champin\inst{1,2} \and Lionel M\'edini\inst{1} \and Fr\'ed\'erique Laforest\inst{1}}
\authorrunning{J. Bruyat et al.}
\institute{Université de Lyon, INSALyon, UCBL, LIRIS CNRS UMR 5205, Lyon, France \email{\{firstname.lastname\}@liris.cnrs.fr} \and
    W3C / ERCIM, Sophia Antipolis, France}

\maketitle

\begin{abstract}
Converting property graphs to RDF graphs allows to enhance the interoperability of knowledge graphs.
But existing tools perform the same conversion for every graph, regardless of its content. In this paper, we propose PREC, a user-configured conversion of property graphs to RDF graphs to better capture the semantics of the content.

\end{abstract}

\section{Introduction}

Property graphs (PGs) are a family of graph models where nodes and edges can have properties. They are widely used in industry. RDF is a model where each node and relation has only one label. RDF provides better syntactic and semantic interoperability, being a W3C standard.

In this paper, we describe our PG-to-RDF Experimental Converter (PREC)\footnote{\url{https://github.com/BruJu/PREC}} that aims to convert PGs to RDF while semantically enriching the contained data. Unlike other approaches, PREC allows users to explicitly specify this enrichment.
We first remind what PGs and RDF graphs are, and review some of the existing tools for converting graphs from one model to the other. We then describe the translation process of PREC. We finally discuss the proposed solution.\freefootnote{\textit{Acknowledgement:} This work was partly supported by the EC within the H2020 Program under grant agreement 825333 (MOSAICrOWN).}\freefootnote{Copyright © 2021 for this paper by its authors. Use permitted
under Creative Commons License Attribution 4.0 International (CC BY 4.0).}

\section{Definitions and state of the art}

In this section, after introducing property graphs and RDF graphs, we review some of the work that already exists to bridge the gap between them.

\subsection{Different types of graphs}

\subsubsection{Property graphs}

Property graphs are not a unique model: each vendor (Neo4j, Amazon Neptune, OrientDB\dots) proposes its own set of features for its property graph model. 
We focus on a model covering a large majority of PG implementations: in this paper we define a PG as a graph with nodes and edges where each edge has a source node, a destination node and one label. A label is a string that describes the relation between the two linked nodes. Nodes can have any number of labels, including none. Nodes and edges can also have properties: a list of key-value pairs, where the key is a string and the value can be a string, a number or a list thereof. Properties can also have properties, called meta-properties.

\subsubsection{Resource Description Framework graphs (RDF)}

The RDF graph model \cite{Wood:14:RCA} is a W3C standard. An RDF graph is defined as a set of $(subject,\allowbreak predicate,\allowbreak object)$ triples. The terms composing a triple can either be IRIs (Internationalized Resource Identifier), blank nodes (except for predicates) or literals (for objects only).

Unlike PGs, RDF has a default semantic: each triple of the RDF graph is deemed true. Removing any triple does not change the truth value of other triples (any RDF graph must entail any subset of itself). Moreover, IRIs are assumed to have a shared semantics, defined by their issuer.

\subsection{Existing syntax}

For an end user, the most notable difference between the RDF model and the PG model is the ability of the latter to add properties to the edges of the graph. Since the creation of RDF, many methods to add annotations to triples have been proposed. The standard RDF reification~\cite{Manola:04:RP} consists in creating a node that represents the triple, and is linked to the subject, predicate and object with \texttt{rdf:subject}, \texttt{rdf:predicate} and \texttt{rdf:object}. Other common methods include singleton properties~\cite{nguyen2014don} or even using named graphs. \textit{A tale of two graphs}~\cite{das2014tale} explores these different methods to translate PG edge properties to RDF. Hartig and Thompson proposed RDF-star~\cite{DBLP:journals/corr/HartigT14}: an extension of the RDF model to allow the use of triples as the subject or the object of other triples, which emulates the ability of PGs to add properties to edges.

\subsection{Existing converters}

Most of the existing work studies how to expose an RDF Graph with a PG API. G2GML~\cite{chiba2020g2gml} uses SPARQL queries to populate a PG. rdf2neo~\cite{brandizi2018getting} populates a PG to have an API that the authors consider easier to use than the usual RDF APIs. Gremlinator~\cite{thakkar2018two} exposes a Gremlin endpoint to query an RDF graph. 

To the best of our knowledge, only two systems allow to convert from PG to RDF.
Neosemantics\footnote{\url{https://github.com/neo4j-labs/neosemantics}} enables users to convert in both directions. Moreover it aims at enriching the PG model with RDF features, \textit{e.g.} inference and model validation with SHACL. 
Tomaszuk et al.~\cite{tomaszuk2020pgo} developed the property graph ontology (PGO) implemented in the graphConv tool. It provides a proper way to describe PGs by using RDF. The authors proved the reverse transformation is also possible.
But neither Neosemantics nor graphConv are configurable: the same conversion is applied regardless of the information stored in the PG. While NeoSemantics creates a contextual node when populating a PG from an RDF graph to be able to convert back to the same RDF graph, this node is not intended to be created or modified by the user.

\subsection{Mapping languages}

Tools already exist to produce RDF triples from another source. R2RML~\cite{Cyganiak:12:RRR} enables a SQL to RDF conversion through a mapping described by the user. The mapping is described thanks to IRIs that are templated with the SQL table fields. This mapping has been further expanded to other formats, like JSON, with RML~\cite{dimou2014rml}. Another noteworthy tool is JSON-LD~\cite{kellogg2019json}, which lets the user write a context in JSON to produce RDF graphs from a data JSON file.

\section{PREC}

\begin{figure}[t]
\centering
\includegraphics[width=1\textwidth]{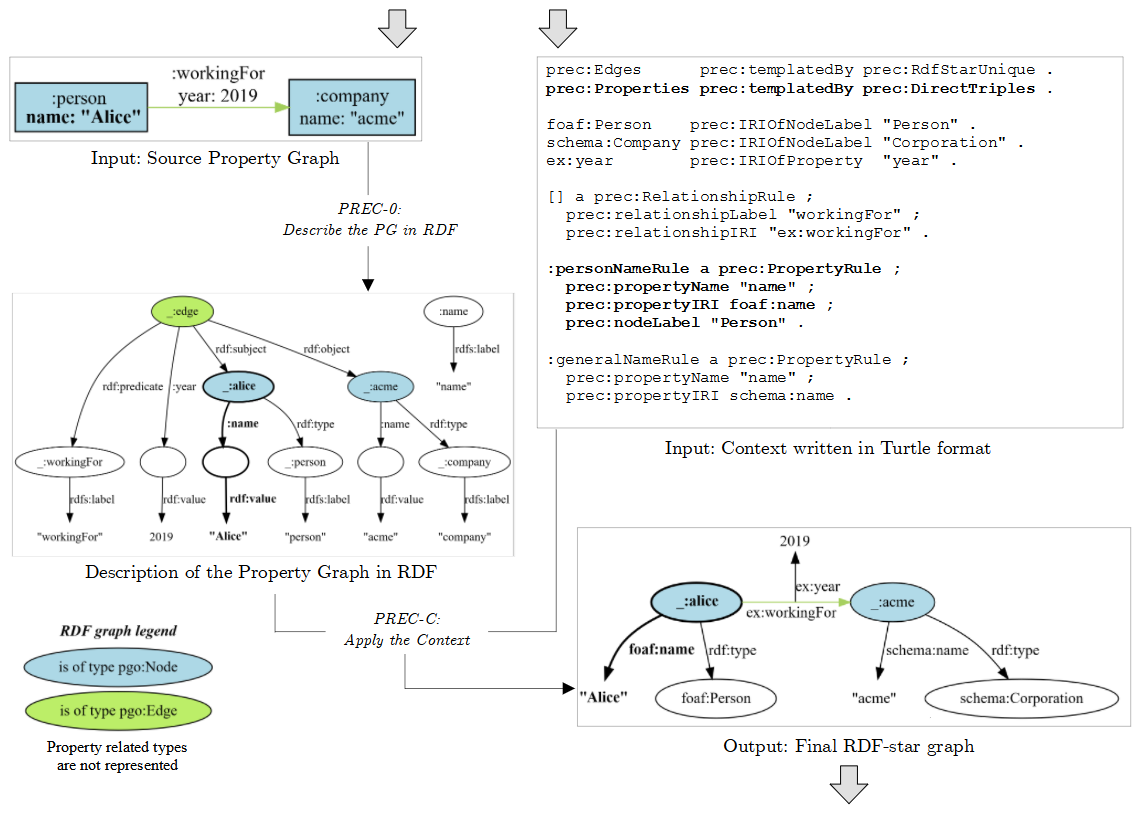}
\caption{General overview of PREC.}
\label{precgeneral}
\end{figure}

In this paper, we propose a user-configured PG-to-RDF translation: instead of using the PG as the sole input, PREC also uses a \textit{context} inspired from JSON-LD. Such a context describes the mapping between the terms used in the PG and IRIs, and templates to represent the different properties and edges in the resulting RDF graph. This enables to better capture and translate the implicit semantics contained in the PG. To do so, PREC performs a two-steps translation as shown in Figure~\ref{precgeneral} and described in the following sections.

\subsection{PREC-0: describing the PG in RDF}

To tame the great diversity of PG models, we propose to work on a uniform graph model. This uniform model maps each PG to an RDF graph that describes its structure, similarly to the ones produced by graphConv~\cite{tomaszuk2020pgo}.

We built our own model of RDF graph that can describe the nodes, edges, labels and properties of any PG. Most terms used in our model are from the RDF and RDFS namespaces. We also use terms from the property graph ontology. Our own PREC namespace defines some extra terms for property typing.
Unlike graphConv, we support meta-properties, and we create a blank node for each label, instead of directly using a literal to represent them. 

The PREC-0 mapping from a PG to its description in RDF is reversible: we can rebuild the nodes and the edges from their descriptions. As a blank node is built to identify every node, edge, label and property of the PG, we ensure that no information is lost in the process.

\subsection{PREC-C: applying the context}

The originality of our work is the ability to apply a context. A context elicits the semantics of the labels and property names of the PG by mapping them to IRIs and describing how to represent them in RDF.

\subsubsection{The context vocabulary}
\hspace{-1ex}\footnote{\url{https://bruy.at/prec}}
A PREC context is an RDF-star graph describing a set of rules. Rules mainly concern node labels, edges and properties.

An example of rule is the \texttt{:personNameRule} shown in the context of Figure~\ref{precgeneral}. This rule is applied to the properties with the key ``name", only for nodes that hold the label ``Person". Each such property is mapped to the IRI \texttt{foaf:name} and directly modelled as a triple. In the same example, the name of Alice, a person, is captured by the rule and transformed. The name of acme, a company, is transformed by the other more general rule \texttt{:generalNameRule}.

\subsubsection{Templating}
\label{modelize}

As our goal is to let users describe how to model the PG components, destination templates are not actually hard coded in PREC-C. Listing~\ref{lst:reversereverse} shows how the \texttt{prec:RdfStarUnique} template (used in the example of Figure~\ref{precgeneral}) is defined\footnote{The pvar namespace contains placeholders for the PG elements matched by the rules.}. In a similar fashion, users can build their own templates.

\begin{lstlisting}[language=turtle, caption=An edge template that maps the PG edge to one RDF triple., label={lst:reversereverse},basicstyle=\ttfamily\scriptsize]
prec:RdfStarUnique a prec:EdgeTemplate ; prec:composedOf
    # The edge as an asserted triple
    <<    pvar:source pvar:edgeIRI pvar:destination               >> ,
    # Keep the information that it was an edge in the PG
    << << pvar:source pvar:edgeIRI pvar:destination >> a pgo:Edge >> ,
    # Assign properties to the embedded triple
    << << pvar:source pvar:edgeIRI pvar:destination >>
            pvar:propertyPredicate pvar:propertyObject >> .
\end{lstlisting}

\subsubsection{Design Choices for the Context Vocabulary}

The context is described in RDF-star. The context vocabulary aims to be agnostic of implementation details, such as the structure of the initial description of the PG in RDF, and the actual implementation of the rules:
\begin{itemize}
    \item Rules in contexts refer to the labels and property names used in the PG. An end user does not need to know the implementation details of PREC-0 to write a context and use PREC.
    \item Contexts are declarative rather than imperative: the order in which rules are declared is not relevant, making contexts easier to write and maintain.
    \item Instead of the declaration order, rules are executed in an order that depends on their type : first meta-property rules, then property rules, then edge rules. It avoids harmful interactions as the production of the latter depends on the production of the former.
\end{itemize}

\section{Preliminary functional evaluation}

In terms of PG-to-RDF graph conversion, PREC provides the same options as the two existing tools.

\paragraph{Neosemantics}

Reproducing the behaviour of Neosemantics to convert Neo4j graphs to RDF is trivial. This can be achieved by keeping the first two lines of the context in Figure~\ref{precgeneral}, \textit{i.e.} modeling every property with the \texttt{prec:\allowbreak{}DirectTriples} template and every edge with the \texttt{prec:RdfStarUnique} template.

\paragraph{PG Ontology}

PREC is also able to produce graphs that follows the PG Ontology~\cite{tomaszuk2020pgo}. To do so, the user needs to provide a context with explicit templates that describe how nodes, edges and properties are represented\footnote{\textit{e.g.} \url{https://gist.github.com/BruJu/21ef88497217da5e2f46b0eea7cf7cac}}.

\section{Challenges and perspectives}

In this paper, we introduced the main motivations behind PREC: a semantic translation from PG to RDF. It is performed through a two-steps conversion: the first step converts a PG to an RDF graph that describes its structure by using a common model. The second one uses a context provided by the user that guides the conversion from an RDF description of the PG structure to an idiomatic RDF graph capturing the implicit semantics of the PG.

We are now considering several perspectives for this work:

\begin{itemize}
    \item \emph{Information loss}: We intend to detect information loss due to the chosen RDF representation of the properties and edges of the PG.
    \item \emph{Reversibility}: Users may want to revert the process, to go back from RDF to the original PG. As the destination template is free, we have to carefully detect which rule each triple of the graph comes from to build back the graph.
    \item \emph{Other RDF transformation tools}: the second step of our approach transforms the RDF graph produced by PREC-0 into another RDF graph, for which existing tools (\textit{e.g.} SPARQL CONSTRUCT) could also be used. We plan to study to which point such existing tools could be reused and refocused to the specific needs of PG translation to RDF.
\end{itemize}


\bibliographystyle{splncs04}
\bibliography{index}

\end{document}